\documentclass[twocolumn,showpacs,preprintnumbers]{revtex4-2}
\usepackage[caption=false]{subfig}
\usepackage{amsmath}  
\usepackage{amsfonts}  
\usepackage{graphicx} 
\usepackage{enumerate}
\usepackage{tikz}
\usetikzlibrary{shapes,arrows, arrows.meta, calc, positioning}
\usepackage[colorlinks=true,linkcolor=blue,urlcolor=blue,citecolor=blue]{hyperref}
\begin{document}
\newcommand{\ds}{\displaystyle}
\newcommand{\be}{\begin{equation}}
\newcommand{\en}{\end{equation}}
\newcommand{\bea}{\begin{eqnarray}}
\newcommand{\ena}{\end{eqnarray}}

\title{On non-equatorial embeddings into $\mathbb{R}^3$ of spherically symmetric wormholes with topological defects}
\author{Mauricio Cataldo}
\altaffiliation{mcataldo@ubiobio.cl} 
\affiliation{Departamento de
F\'\i sica, Universidad del Bío-B\'io, Casilla 5-C,
Concepci\'on, Chile.\\}
\author{Daniel Cuevas}
\altaffiliation{daniel.cuevas1501@alumnos.ubiobio.cl} 
\affiliation{Departamento de
F\'\i sica, Universidad del Bío-B\'io, Casilla 5-C,
Concepci\'on, Chile.\\}

\date{\today} 
\begin{abstract}
Traditionally, the embedding procedure for spherically symmetric spacetimes has been restricted to the equatorial plane $\theta = \pi/2$. This conventional approach, however, encounters a fundamental limitation: not every spherically symmetric geometry admits an isometric embedding of its equatorial slice into three-dimensional Euclidean space. When such embeddings are not possible, the standard geometric intuition becomes inapplicable. In this work, we generalize the embedding procedure to slices with arbitrary polar angles $\theta \neq \pi/2$, thereby extending the visualization and analysis of spacetimes beyond the reach of traditional methods. 

The formalism is applied to Schwarzschild-like wormholes and to a generalized Minkowski spacetime with angular deficit or excess, which are particularly relevant since their equatorial slices cannot be consistently embedded in $\mathbb{R}^3$. In these cases, we identify the explicit constraints on the radial coordinate, polar angle, and geometric parameters required to guarantee consistent embeddings into three-dimensional Euclidean space.

\vspace{0.5cm}
\end{abstract}
\smallskip
\maketitle 
\preprint{APS/123-QED}


\section{Introduction}
The isometric embeddings of two-dimensional slices of spacetime metrics into three-dimensional Euclidean space provide a powerful visual tool for understanding the geometric properties of exact solutions of the Einstein equations. 

From a pedagogical perspective, these spacetime embeddings are essential for illustrating complex concepts in general relativity and differential geometry, enhancing physical and mathematical intuition by transforming abstract concepts into tangible representations. In this way, fundamental aspects such as spacetime curvature, horizons, and singularities can be visualized without requiring advanced mathematical tools \cite{Visser:1996,I.Nikitin:2021,Collas:2011py}. This type of visualization facilitates the understanding of geodesics and the motion of test particles, allowing one to visualize trajectories and gain a clearer comprehension of concepts such as the throat of a wormhole  \cite{T.Muller:2004,D.Weiskopf:2006,O.James:2015}.
Despite these advantages, there are limitations, as it is not always possible to find isometric embeddings in three dimensions for spacetime metrics. This is because some metrics require higher-dimensional spaces for their embedding \cite{M.Janet:1926,T.Cartan:1927,J.Nash:1956}, and certain aspects of spacetime geometry may be entirely non-visualizable, particularly those related to the temporal nature of the spacetime metrics.

Motivated by the above considerations, in this work we extend the study of the embedding formalism by applying it to two-dimensional spacelike slices defined by $t=\text{const.}$ and $\theta=\text{const.}$ in spherically symmetric spacetimes. It is worth noting that, in most investigations of the geometric properties of such spacetimes, the embedding procedure is usually restricted to equatorial slices with $t=\text{const.}$ and $\theta=\pi/2$. Numerous studies have investigated embeddings of equatorial slices in the most extensively analyzed solutions, with the Schwarzschild black hole and wormhole geometries serving as the most prominent examples~\cite{Morris:1988cz1,Collas:2011py,O.James:2015,Hao:2023igi,Sadegh,Kar:2024ctd}.

In the case of the Schwarzschild black hole
\begin{eqnarray*}
ds^2=\left(1-\frac{2M}{r}\right)dt^2 - \frac{dr^2}{1-\frac{2M}{r}} - r^2\left( d\theta^2 + \sin^2 \theta \, d\varphi^2 \right),
\end{eqnarray*}
it is possible to embed the equatorial slice for any value of the mass $M>0$.  

Similarly, for the family of wormholes described by
\begin{eqnarray*}
ds^2 = e^{2\phi(r)}dt^2 - \frac{dr^2}{1-\left(\tfrac{r_0}{r}\right)^\alpha} - r^2\left( d\theta^2 + \sin^2 \theta \, d\varphi^2 \right),
\end{eqnarray*}
the equatorial slice can also be embedded for arbitrary positive values of $r_0$ and $\alpha$. This family includes, as a particular case, the Schwarzschild wormhole solution corresponding to $\alpha = 1$~\cite{Morris:1988cz1}.

The justification commonly given for restricting the analysis to the equatorial slice relies on several compelling arguments. First, spherical symmetry guarantees that the equatorial plane ($\theta = \pi/2$) encodes all essential information about the radial curvature structure, which is the most physically relevant aspect of these spacetimes. Second, this dimensional reduction from a four-dimensional spacetime to an effective two-dimensional problem makes the embedding procedure technically tractable while preserving the key geometric features.

However, this conventional understanding encounters a fundamental limitation that has received insufficient attention in the literature. Not all spherically symmetric spacetimes admit isometric embeddings of their equatorial slices into three-dimensional Euclidean space. When such embeddings fail to exist the standard geometric intuition breaks down entirely. This raises a profound question with both theoretical and practical implications: How can we develop systematic methods to analyze, visualize, and understand the geometric properties of spherically symmetric spacetimes whose equatorial slices cannot be embedded in flat three-dimensional space?

To systematically investigate this fundamental limitation, we examine the Schwarzschild-like wormholes~\cite{Cataldo-Liempi-Rodriguez:2017}. This solution provides an ideal theoretical framework for our investigation because it exhibits the precise pathology we seek to understand, while maintaining the familiar spherical symmetry that normally guarantees successful embeddings; nevertheless, within certain parameter regimes its equatorial slices become non-embeddable in three-dimensional Euclidean space $\mathbb{R}^3$.

The geometric obstruction arises from topological features inherent to the wormhole structure, specifically the presence of conical singularities that violate the necessary conditions for isometric embedding. This makes the Schwarzschild-like wormhole solution particularly valuable as a concrete example where traditional visualization methods fail, thereby forcing us to confront the fundamental question of how to analyze such geometries when conventional tools prove inadequate. In what follows, we provide a concise review of the essential features of this solution, with particular emphasis on the geometric properties that give rise to embedding limitations.

The Schwarzschild-like wormholes constitute a solution that slightly generalizes the Schwarzschild wormhole~\cite{Morris:1988cz1}. This solution is obtained within the Morris and Thorne framework by imposing a linear behavior on the radial coordinate in the shape function of the metric
\begin{eqnarray}
ds^2=e^{2\Phi(r)}dt^2-\frac{dr^2}{1-\frac{b\left(r\right)}{r}}-r^2\left(d\theta^2+\sin^2\theta d\phi^2\right), \,\,\, \label{generalmetric}
\end{eqnarray}
where $\Phi(r)$ and $b(r)$ are the redshift and shape functions, respectively. It is noteworthy that the metric~(\ref{generalmetric}) represents the general form of any static, spherically symmetric gravitational field. The authors constructed a wormhole spacetime by imposing specific conditions on $\Phi(r)$ and $b(r)$ to ensure the existence of a bridge and to guarantee that the wormhole is traversable for a human traveler, while simultaneously avoiding event horizons and singularities~\cite{Morris:1988cz1,Morris:1988cz2}. A wide variety of Morris and Thorne wormhole solutions can be constructed by modifying the redshift and shape functions in the metric~(\ref{generalmetric}). These solutions arise both within the framework of general relativity and in the context of modified gravity theories.

In Ref.~\cite{Cataldo-Liempi-Rodriguez:2017}, the wormhole solution under consideration is obtained by expressing the shape function in the form $b(r) = \left(1-\beta\right)r_0 + \beta r$, where $\beta$ and $r_0$ are constant parameters with $r_0 > 0$. Subsequently, the metric~(\ref{generalmetric}) takes the form
\begin{eqnarray}
ds^2 &=& e^{2\Phi(r)} dt^2 - \frac{dr^2}{\left(1-\beta\right)\left(1-\frac{r_0}{r}\right)} \nonumber \\
&& - r^2 \left(d\theta^2 + \sin^2\theta d\phi^2\right). \label{schwarzschild_like_wormholes}
\end{eqnarray}

The redshift function $\Phi(r)$ must remain finite throughout spacetime to ensure the absence of horizons and singularities, thereby achieving a traversable wormhole. Since the metric of the Schwarzschild-like wormholes is expressed in spherical coordinates, it is evident that the angular part of the metric represents the geometry of a two-dimensional sphere with radius $r$. The angular coordinates $\theta$ and $\phi$ vary from $0$ to $\pi$ and $0$ to $2\pi$, respectively.

Let us now clarify the implications of the $\beta$ parameter by examining specific cases of the metric~(\ref{schwarzschild_like_wormholes}). In particular, for solutions with a vanishing redshift function $\Phi(r)$ at spatial infinity, or for wormholes with vanishing tidal forces, Eq.~(\ref{schwarzschild_like_wormholes}) yields
\begin{eqnarray}
ds^2 = dt^2 - \frac{dr^2}{1-\beta} - r^2 \left(d\theta^2 + \sin^2\theta \, d\phi^2\right),
\label{sssss}
\end{eqnarray}
in the limit $r \rightarrow \infty$. It follows that the Lorentzian signature of the metric requires $\beta < 1$.

It is worth mentioning that the presence of the $\beta$-parameter modifies the Minkowski metric by introducing topological defects into an otherwise flat spacetime. Notice that Morris and Thorne specifically analyzed asymptotically flat wormholes~\cite{Morris:1988cz1,Morris:1988cz2}, which connect two regions either within the same universe or between different universes. For this class of wormholes, the condition $b(r)/r \to 0$ as $r \to \infty$ is imposed. By contrast, for Schwarzschild-like wormholes one obtains $b(r)/r \to \beta$ as $r \to \infty$, indicating that the presence of the $\beta$ parameter makes the spacetime asymptotically conical rather than asymptotically flat.

Note that, for the metric~(\ref{sssss}), the area of a sphere with radius $r$ is locally $4 \pi r^2$. However, despite its relatively simple structure, the Ricci scalar is given by $R = 2\beta/r^2$, indicating that the metric is not locally flat for $\beta \neq 0$.

It can be shown that the metric~(\ref{sssss}), with the $\beta$-parameter in the range $0<\beta<1$, may be associated with a global monopole, which is a type of topological defect that can arise in certain theories of cosmology and particle physics, particularly in models where a global symmetry is spontaneously broken~\cite{Barriola}.
To compare the above metric directly with the metric of a global monopole, we rescale the \( r \) coordinate and rewrite the metric~(\ref{sssss}) as
\begin{eqnarray}
ds^2 = dt^2 - dr^2 - (1-\beta) r^2 \left(d\theta^2 + \sin^2\theta \, d\phi^2\right).
\label{tsssss}
\end{eqnarray}
This allows us to identify \( \beta = 8 \pi G \eta^2 \)~\cite{Barriola}. Since \( 8 \pi G \eta^2 > 0 \), the metric of a global monopole describes a spacetime with a deficit solid angle.

However, notice that the metric~(\ref{tsssss}) has a more general character. In this case, the solid angle and the area of a sphere with radius $r$ are not $4\pi$ and $4 \pi r^2$, respectively, but $4 \pi(1-\beta)$ and $4 \pi (1-\beta)r^2$. So if $0<\beta<1$ the metric~(\ref{sssss}) describes a space with a deficit solid angle, and for $\beta<0$ this metric includes spaces with an excess of solid angle. As expected, the metric~(\ref{tsssss}) is not locally flat, as indicated by the Ricci scalar curvature $R = \frac{2\beta}{(1-\beta)r^2}$.

It is important to emphasize that, for the metric~(\ref{sssss}) with $\beta \neq 0$, the embedding of an equatorial slice into $\mathbb{R}^3$ is strictly possible only in the presence of an angular deficit.
Conversely, whenever this metric exhibits an angular excess, such an embedding becomes strictly forbidden, and the equatorial plane is thereby non-embeddable in $\mathbb{R}^3$.

Based on this crucial fact, in the present work we extend the standard isometric embedding procedure, typically applied to the equatorial plane $\theta = \pi/2$, to arbitrary angles $\theta \neq \pi/2$. This generalization allows us to explore the properties of two-dimensional spatial slices of spherically symmetric metrics whose equatorial slices cannot be embedded into $\mathbb{R}^3$, thereby providing valuable insights into the geometry under study.

In this way, the application of the generalized isometric embedding procedure offers deeper insight into the metric~(\ref{schwarzschild_like_wormholes}), showing how the presence of the parameter $\beta$ affects the embedding of specific slices and, in contrast to the Schwarzschild wormhole, implies that the embedding of the equatorial plane does not always exist. In particular, we present a detailed analysis of the background metric~(\ref{sssss}), emphasizing that for $\beta < 0$ the equatorial slice cannot be embedded into $\mathbb{R}^3$.

This paper is organized as follows: In Sec.~II, we briefly outline the key aspects of the extension of the standard isometric embedding procedure for the equatorial plane \( \theta = \pi/2 \) of a spherically symmetric metric to other angles, thereby enabling the embedding of slices with \( \theta \neq \pi/2 \). Section~III focuses on the embeddings of specific slices of Schwarzschild-like wormholes that exhibit conical singularities. In Sec.~IV, we analyze the embeddings of slices of Schwarzschild-like wormholes that do not exhibit conical singularities. Finally, our conclusions are presented in Sec.~V.

\section{Extension of the standard embedding procedure}\label{SectionII}
The standard procedure for embedding two-dimensional slices into three-dimensional Euclidean space begins with the isometric embedding method, where one considers a slice with $t=\text{const.}$ and $\theta=\pi/2$ of a generic spherically symmetric metric~(\ref{generalmetric}). Owing to the generality of this metric, the method applies to a wide class of spacetimes, allowing for a unified description of their embedding properties. In particular, it provides a means to visualize the shape and size of the equatorial slice, which captures the key geometric features of spherically symmetric spacetimes. This procedure has been discussed in detail in several works, including Refs.~\cite{Collas:2011py,O.James:2015,T.Muller:2004,Cataldo-Liempi-Rodriguez:2017,Rueda2022,Errehymy2024,Blázquez-Salcedo2021}.

In this section, we extend this procedure to include slices with $\theta \neq \pi/2$ in the analysis. Specifically, we consider two-dimensional spacelike slices with $t = \text{const}$ and $\theta = \text{const}$ of the metric~(\ref{generalmetric}), embedding them into ordinary three-dimensional Euclidean space expressed in cylindrical coordinates as
\begin{equation}
ds_{\text{cyl}}^2 = d\rho^2 + \rho^2 d\phi^2 + dz^2. \label{cylindrical}
\end{equation}

For the embedding construction, we work with the metric~(\ref{generalmetric}) in Schwarzschild coordinates, characterized by the functions $\Phi(r)$ and $b(r)$ that encode the specific properties of the spacetime. Since this representation covers the entire class of static, spherically symmetric geometries, the extension of the embedding procedure applies in full generality.

Therefore, by imposing the constraints $t=\text{const}$ and $\theta=\text{const}$ on the general metric~(\ref{generalmetric}), we extract the two-dimensional spatial slice
\begin{eqnarray}
ds_{\text{slice}}^2=\frac{dr^2}{1-\frac{b\left(r\right)}{r}}+r^2\sin^2\theta \, d\phi^2, \label{slicegeneralmetric}
\end{eqnarray}
where we require $\theta \neq 0$ and $\theta \neq \pi$ to ensure that the slice~(\ref{slicegeneralmetric}) remains non-degenerate and preserves its two-dimensional character. This constraint is essential because at the poles ($\theta = 0, \pi$), the angular coordinate $\phi$ becomes undefined, causing the metric to collapse to a one-dimensional line element.

To find the embedding, we identify the components of the metric slice~(\ref{slicegeneralmetric}) with the corresponding components of the cylindrical coordinates~(\ref{cylindrical}). By comparing these metrics we may identify 
\begin{eqnarray}
\rho=r \,\sin \, \theta. 
\label{rhotheta} 
\end{eqnarray}
The remaining parts of both metrics should match, so 
\begin{eqnarray*}
\frac{dr^2}{1-\frac{b\left(r\right)}{r}}=\sin^2  \, \theta dr^2+dz^2,
\end{eqnarray*}
obtaining finally
\begin{eqnarray}
\frac{dz(r)}{dr}=\pm \left(\frac{1}{1-\frac{b(r)}{r}}-\sin^2 \, \theta \right)^{\frac{1}{2}},
\label{z(r)} 
\end{eqnarray}
where, for any given constant value of $\theta$, the constraint $0 < \sin^2 \theta \leq 1$ is satisfied. Thus, Eq.~(\ref{z(r)}) determines how the spatial slice~(\ref{slicegeneralmetric}) is embedded in flat Euclidean space, with the function $z(r)$ reflecting the curvature of the original spacetime.

The necessary and sufficient condition for the square root in~(\ref{z(r)}) to remain real is 
\begin{equation*}
1-\frac{b(r)}{r} \leq \frac{1}{\sin^2\theta}.
\end{equation*}
From this relation, one can directly derive the inequalities that constrain the radial coordinate $r$, the polar angle $\theta$, or the parameters of the shape function $b(r)$.

This generalization provides a systematic method to analyze cases in which the equatorial embedding fails, offering new insights into the geometry of the two-dimensional slices. Moreover, this procedure has a broad applicability: it can in principle be employed to study embeddings of non-equatorial slices in other static spacetimes, including black holes, wormholes, and other spherically symmetric gravitational configurations.

\section{Embeddings of spaces with topological defects}\label{SectionIII}
We are now prepared to analyze the deviations introduced by the parameter $\beta$ in the schwarzschild-like metric~(\ref{schwarzschild_like_wormholes}), relative to the standard Schwarzschild wormhole, whose equatorial slices can all be embedded in $\mathbb{R}^3$, as well as those induced by the same parameter in the metric~(\ref{sssss}), relative to flat spacetime.

As we shall see, this approach makes it possible to identify precise geometric conditions on the angular coordinate $\theta$, the radial coordinate $r$, and the parameter $\beta$ that govern the existence of embeddings in $\mathbb{R}^3$, thereby extending the scope of embedding diagrams beyond the traditional equatorial plane.

For the Schwarzschild-like wormholes~(\ref{schwarzschild_like_wormholes}), Eq.~(\ref{z(r)}) can be recast in the form
\begin{eqnarray}
z(r) = \pm \int \left[\frac{1}{\left(1-\beta\right)\left(1-\tfrac{r_0}{r}\right)} - \sin^2 \theta \right]^{1/2} \, dr + C^{\pm}, \,\,\,\,\,\,
\label{conicasssssG}
\end{eqnarray}
where $C^{\pm}$ denote constants of integration. This representation will allow us to determine the conditions under which the embedding remains well-defined and to characterize the influence of $\beta$ on the global geometry of the slices.

To begin, we will analyze the embeddings of two-dimensional slices of the metric~(\ref{sssss}), described by
\begin{eqnarray}
ds_{slice}^2=\frac{dr^2}{1- \beta}+r^2\sin^2\theta \, d\phi^2, \label{conicalslice}
\end{eqnarray}
In order to do this we put $r_0=0$ into Eq.~(\ref{conicasssssG}), obtaining
\begin{eqnarray}
z(r) = \pm \left(\frac{1}{1 - \beta}-\sin^2 \theta \right)^{\frac{1}{2}} \, r +C^{\pm}.
\label{conicasssss}
\end{eqnarray}
In general, this equation represents a tilted cone with its vertex shifted along the $z$-axis. When $C^{\pm} = 0$ this equation reduces to a cone whose vertex is located at the origin. 

It is crucial to highlight that Eq.~(\ref{conicasssss}) represents a cone as $z = z(r)$, instead of $z = z(\rho)$, which is the appropriate form in cylindrical coordinates. Consequently, to obtain the correct equation for embedded cones in cylindrical coordinates, it is necessary to also consider Eq.~(\ref{rhotheta}). Thus, combining Eqs.~(\ref{conicasssss}) and~(\ref{rhotheta}), we obtain
\begin{eqnarray}
z(\rho) = \pm \frac{\left( \frac{1}{1 - \beta} - \sin^2\theta \right)^{\frac{1}{2}}}{\sin\theta} \,  \rho,
\label{zderho}
\end{eqnarray}
where we have set \(C^{\pm} = 0\). Notice that, due to the cylindrical symmetry, the considered embeddings are surfaces of revolution. In this way, the procedure enabled us to determine the parametric equation of the slice \(\theta = \text{const}\) being embedded into the ordinary three-dimensional Euclidean space.

An interesting particular case to consider is \(\beta = 0\), which represents flat spacetime. Substituting \(\beta = 0\) into Eq.~(\ref{zderho}), we obtain
\begin{eqnarray}
z(\rho) = \frac{\rho}{\tan \theta}.
\end{eqnarray}
In this case, $\theta$ is the opening angle of the conical sections with respect to the $z$-axis. It must be noted that this equation represents the embedding of any conical slice with $t = \text{const}$ and $\theta = \text{const}$ of the flat spacetime written in spherical coordinates. In this way, we can conclude that the \(\beta\)-parameter in Eq.~(\ref{zderho}) introduces changes to the geometric structure of flat spacetime.

Now, we can determine the opening angle of the conical slices with respect to the \(z\)-axis for $\beta \neq 0$ by using Eq.~(\ref{zderho}). By denoting this angle as \(\alpha\), we obtain
\begin{eqnarray}
\alpha = \arctan \left(\pm \frac{\sin \theta}{ \left(\frac{1}{1 - \beta} - \sin^2 \theta \right)^{\frac{1}{2}}} \right).
\label{anguloalpha}
\end{eqnarray}
It becomes clear that for $\beta=0$ the opening angle $\alpha$ becomes angle $\theta$.

Let us now turn our attention to the conditions required for the existence of embeddings of the metric~(\ref{conicalslice}). For the embedding to exist, Eq.~(\ref{zderho}) dictates that a $\theta$-slice and the parameter $\beta$ must satisfy the following condition:
\begin{eqnarray}
\frac{1}{1 - \beta} - \sin^2 \theta \geq 0.
\label{inequation0}
\end{eqnarray}
In general, the two-parameter inequality~(\ref{inequation0}) allows us 
to fix a particular $\theta$-slice and subsequently restrict 
the range of $\beta$, or to fix a specific value of 
$\beta$ and limit the possible $\theta$-slices.

When a value for $\theta$ is fixed, $\beta$ 
must satisfy the condition
\begin{eqnarray}
1-\frac{1}{\sin^2\theta} \leq \beta<1.
\label{inequation}
\end{eqnarray}
Since, in general, $0 \leq \sin^2 \theta \leq 1$, it follows that $\beta$ can take negative values, lying within the range $-\infty < \beta < 1$. In this way, Eq.~(\ref{inequation}) determines the minimum value that the parameter $\beta$ can take for a given angle $\theta$. Consequently, this minimum value depends on the slice being considered.

On the other hand, by fixing the parameter $\beta$, we define a specific form of the metric given by Equation~(\ref{sssss}). Consequently, the inequality~(\ref{inequation0}) imposes restrictions on the possible $\theta$-slices that can be embedded into three-dimensional Euclidean space.

For $\beta > 0$, the condition~(\ref{inequation0}) does not impose any restrictions on the slices, allowing the embedding of any slice with \(0 < \theta < \pi\) into the ordinary three-dimensional Euclidean space. 

On the contrary, for $\beta<0$, the slices do not 
exist for all values of $\theta$. The condition~(\ref{inequation0}) implies that the following constraints must 
be satisfied:
\begin{eqnarray}
0<\theta \leq \arcsin \left(\sqrt{\frac{1}{1-\beta}}\right), \label{Cbeta0} \\
\pi-\arcsin \left(\sqrt{\frac{1}{1-\beta}}\right) \leq \theta<\pi.
\label{Cbeta1}
\end{eqnarray}
Therefore, embeddings of $\theta$-slices with
\begin{eqnarray}
\arcsin \left(\sqrt{\frac{1}{1-\beta}}\right)<\theta<\pi-\arcsin \left(\sqrt{\frac{1}{1-\beta}}\right)
\label{noposible}
\end{eqnarray}
are not possible in the Euclidean three-dimensional space for the metric~(\ref{conicalslice}) with negative values of $\beta$.

It is worth noting that, the embedding $z(\rho) = 0$ holds a special significance. In this particular case, the constraint~(\ref{inequation0}) leads to the relation $\frac{1}{1 - \beta} = \sin^2 \theta$, which is satisfied only when $\beta \leq 0$. The value $\beta = 0$ corresponds to $\theta = \pi/2$, resulting in the equatorial slice of Minkowski space, which is flat. However, it is worth noting that for negative values of the $\beta$-parameter, there exists a family of slices with $\theta \neq \pi/2$, each corresponding to a specific value of $\beta$, which are flat since their embedding is also given by $z(\rho) = 0$. This unexpected result can be understood by analyzing the metric~(\ref{conicalslice}). By substituting the relation $\frac{1}{1-\beta} = \sin^2 \theta$ into the metric~(\ref{conicalslice}), we obtain
\begin{eqnarray}
ds_{\text{slice}}^2 = \frac{1}{1 - \beta} \left(dr^2 + r^2 \, d\phi^2 \right). \label{conformalslices}
\end{eqnarray}
This metric is conformal to the Euclidean two-dimensional space and, consequently, it is also flat, with the embedding function given by $z(\rho) = 0$.

On the contrary, for the case \(\frac{1}{1-\beta} \neq \sin^2 \theta\), it follows from Eqs.~(\ref{Cbeta0}) and~(\ref{Cbeta1}) that the following inequalities must hold: $0<\theta < \arcsin \left(\sqrt{\frac{1}{1-\beta}}\right)$ and 
$\pi-\arcsin \left(\sqrt{\frac{1}{1-\beta}}\right) < \theta<\pi$.
In this scenario, the metric~(\ref{conicalslice}) can be expressed in the form  
\begin{eqnarray}
ds_{\text{slice}}^2 = \frac{1}{1 - \beta} \left(dr^2 + (1-\beta)r^2 \sin^2 \theta \, d\phi^2 \right). \label{conicalconformalslices}
\end{eqnarray}  
From this, we conclude that the metric is conformal to the conical slice $t = \text{const}$ and $\theta = \text{const}$ of the Minkowski metric, with the angular component modified by the factor $1-\beta$. In this way, we can define a new angle $\tilde{\alpha}$ associated with the metric
\begin{eqnarray*}
ds_{\text{slice}}^2 = dr^2 + r^2 \sin^2 \tilde{\alpha} \, d\phi^2,
\end{eqnarray*}  
from which it follows that $(1-\beta) \sin^2 \theta = \sin^2 \tilde{\alpha}$, and therefore  
\begin{eqnarray}
\tilde{\alpha} = \arcsin\left(\pm \sqrt{(1-\beta)\sin^2 \theta}\right).
\end{eqnarray}  
This expression is equivalent to~(\ref{anguloalpha}), thus defining the opening angle $\tilde{\alpha}=\alpha$ of the cones with respect to the $z$-axis.

We also examine the special case where $z(\rho) = \rho$. Under this condition, Eq.~(\ref{zderho}) yields the following relation:
\begin{equation}
\beta = 1-\frac{1}{2\sin^2\theta}.
\label{diagonal}
\end{equation}
This expression indicates that such a slice can exist only for $\beta \leq 1/2$.

\subsection{Slices with $\theta=\pi/2$}
In general, it is possible to generate meaningful two-dimensional 
slices of the metric (\ref{generalmetric}) for constant $t$ and constant $\theta$, 
with the exception of $\theta=0$ and $\theta=\pi$. For these values of $\theta$, the metric 
becomes degenerate, and we do not obtain a genuine surface. Thus, in 
the analysis that follows, we will consider the polar angle to lie 
within the range $0<\theta<\pi$.

Typically, an embedding diagram of spacetime slices of metric~(\ref{generalmetric}) can be constructed by considering any constant value for $\theta$. However, when constructing embedding diagrams of spherically symmetric slices, the constant value $\theta = \pi/2$ is often used in the literature \cite{T.Muller:2004,D.Weiskopf:2006,M.Hannam:2008,A.Mishra:2018}.  We begin our analysis of embeddings with this case.

The equatorial plane is frequently utilized for embedding slices of spherically symmetric spacetimes because of its simplicity and inherent symmetry. Selecting $\theta = \pi/2$ typically simplifies the equations involved, making the analysis more straightforward and reducing the complexity of the resulting expressions. This is because the equatorial plane represents a plane of reflection symmetry in spherically symmetric spacetimes, where certain angular dependencies, such as those of the $\theta$-coordinate, can be eliminated.

The equatorial plane of the spherically symmetric metric~(\ref{generalmetric}) has the maximal circumference and area for a given radius $r$, making it a convenient and representative slice to study. In such a way, this standard practice simplifies the two-dimensional metric, and allows an ease visualization of the considered slice, retaining in general all the essential features of the spacetime while allowing for a simpler, two-dimensional representation.

As stated above, the constraint~(\ref{inequation}) must be fulfilled for any slice embedding of the metric~(\ref{sssss}). Therefore, embeddings for a slice with $\theta = \pi/2$ exist only within the range $0 \leq \beta < 1$, meaning that the embedding into the ordinary three-dimensional Euclidean space of the equatorial plane of the metric~(\ref{sssss})  is possible only for positive values of $\beta$.

This can also be seen directly by substituting $\theta = \pi/2$ into Eq.~(\ref{zderho}), which yields $z(\rho) = \pm \sqrt{\beta/(1-\beta)}\, \rho$. From this expression, we conclude that if $\beta < 0$, the function $z(\rho)$ becomes imaginary. Thus, it follows that for negative values of $\beta$, it is not possible to embed the equatorial slice of the metric~(\ref{sssss}) in Euclidean space. However, as we shall demonstrate below, it is possible to embed slices of the spacetime~(\ref{sssss}) with $\beta < 0$ and $\theta \neq \pi/2$.

Notice that for the slice with $\theta=\pi/2$, the angle of the cones is given by 
\begin{eqnarray}
\alpha = \arctan\left(\pm \sqrt{(1-\beta)/\beta}\right).
\label{angulo}
\end{eqnarray}
Since $0 < \beta < 1$, we conclude that for a given $\beta$, the angle satisfies $0 < \alpha < \pi/2$ for the positive branch of~(\ref{angulo}).

\begin{figure}
\centering
\includegraphics[scale=0.3]{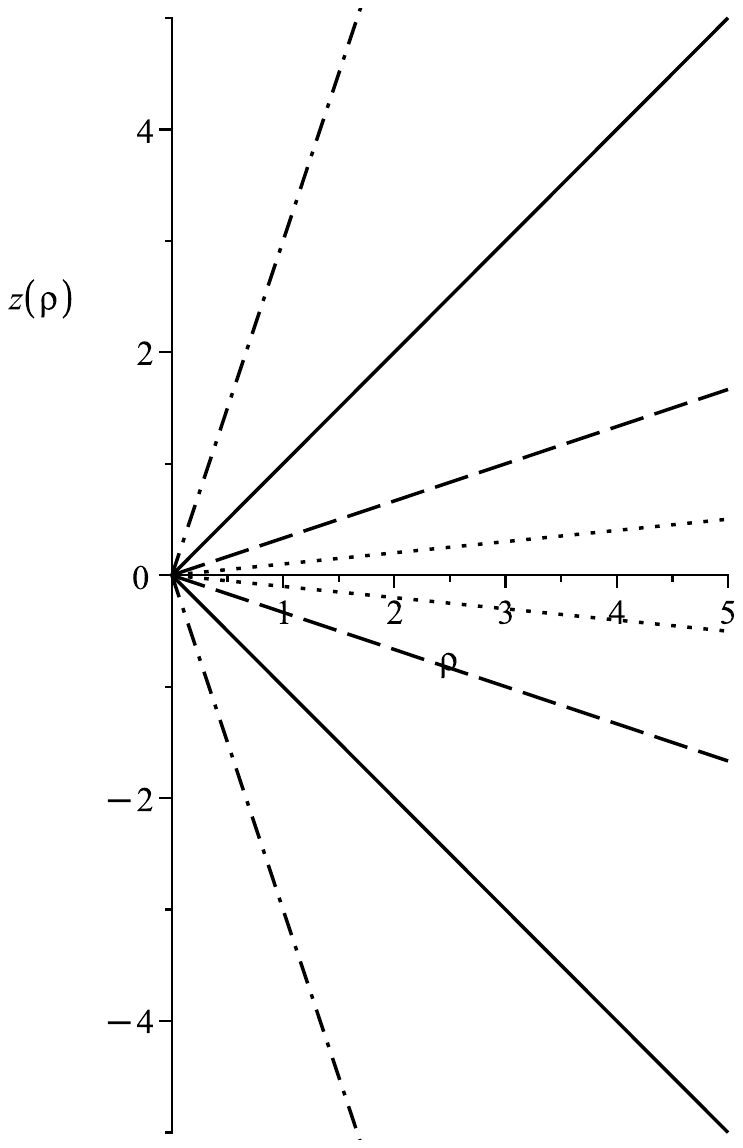}
\caption{The figure shows embedding diagrams of the metric~(\ref{sssss}) derived from Eq.~(\ref{zderho}), with \(\theta = \pi/2\). Equatorial slices are plotted for \(\beta = 9/10\) (dash-dotted line), \(\beta = 1/2\) (solid line), \(\beta = 1/10\) (dashed line), and \(\beta = 1/100\) (dotted line). The plotted slices exhibit a conical singularity.}
\label{cconicaldeficit}
\end{figure}

In Fig.~\ref{cconicaldeficit}, embedding diagrams of equatorial slices of the metric~(\ref{sssss}) are plotted for $\beta = 9/10$, $\beta = 1/2$, $\beta = 1/10$, and $\beta = 1/100$. As expected, for $0 < \beta < 1$, all equatorial slices exhibit a conical singularity. As $\beta \rightarrow 0$, the cones flatten progressively, becoming a plane for $\beta = 0$ (equatorial plane).

\subsection{Slices with $\theta \neq \pi/2$}
The above analysis clearly shows that the embedding of an equatorial slice is not feasible when $\beta$ is negative. Therefore, to embed slices of the metric~(\ref{sssss}) with $\beta < 0$ into the ordinary three-dimensional Euclidean space, we must consider the angular dependencies with $\theta \neq \pi/2$. 

\begin{figure}
\centering
\includegraphics[scale=0.3]{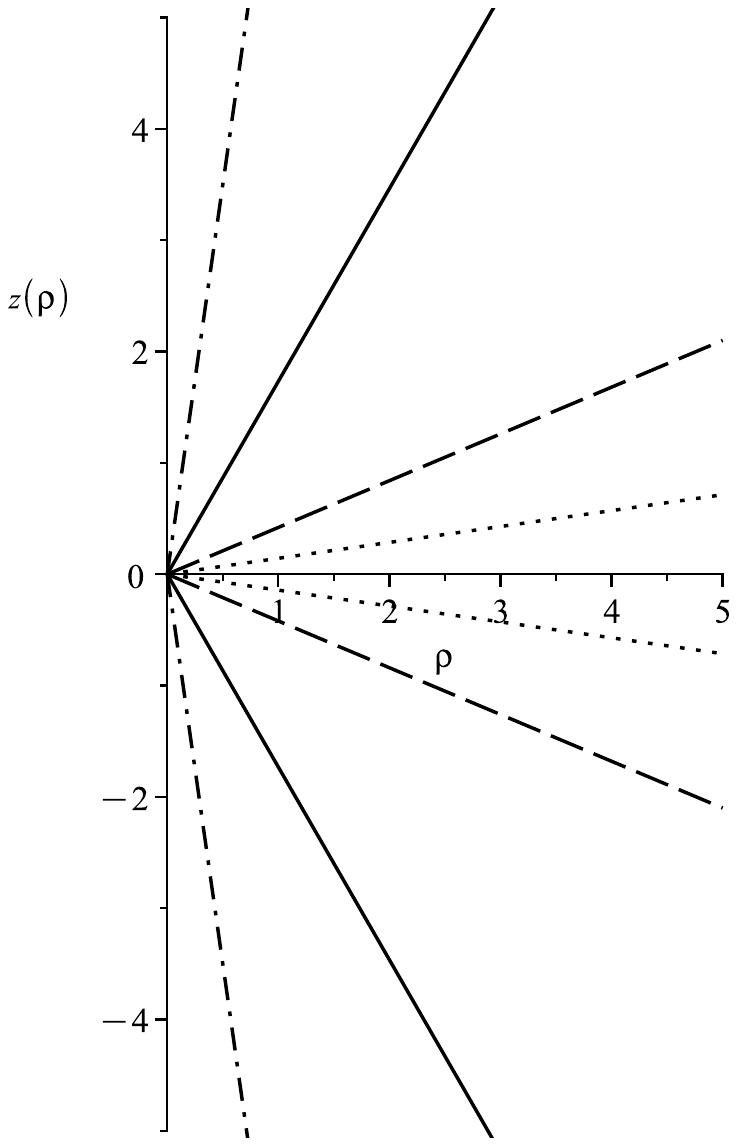}
\caption{The figure shows embedding diagrams of the metric~(\ref{sssss}) derived from Eq.~(\ref{zderho}), with \(\theta = \pi/4\). The diagrams are plotted for \(\beta = 96/100\) (dash-dotted line), \(\beta = 1/2\) (solid line), \(\beta = -7/10\) (dashed line), and \(\beta = -96/100\) (dotted line). All slices exhibit a conical singularity.}
\label{cconicalexceso}
\end{figure}

For visualization, we present the embeddings for selected values of $\beta$, focusing on the slice $\theta=\pi/4$. In this case the Eq.~(\ref{inequation}) gives the range $-1\leq \beta<1$, which allows us to consider negative values for the $\beta$-parameter. In Fig.~\ref{cconicalexceso}, we show slices for $\beta = 96/100$, $\beta =1/2$, $\beta =-7/10$ and $\beta =-96/100$. We conclude that all slices exhibit conical singularities. Notice that now the cones flatten more and more for $\beta \rightarrow -1$.

For $\theta=\pi/4$ the opening angle of the cones~(\ref{zderho}) with respect to the $z$-axis is given by $\alpha = \arctan\left(\sqrt{\frac{1-\beta}{1+\beta}}\right)$. Since in this case $-1 \leq \beta < 1$, it follows that \(0 < \alpha < \pi/2\). For all cones with $0 \leq \beta < 1$ we have $0 < \alpha \leq \pi/4$, while for all cones with $-1 \leq \beta <0$ we have  $\pi/4<\alpha < \pi/2$.

Until now, we have focused on specific $\theta$-slices for different values of the parameter $\beta$. Next, we will fix a specific value of $\beta$ and examine various slices of the chosen spacetime. As discussed earlier, any slice of the metric~(\ref{sssss}) with positive $\beta$ can be embedded into the ordinary three-dimensional Euclidean space. For this reason, we will now restrict our attention to negative values of $\beta$. 

\begin{figure}
\centering
\includegraphics[scale=0.3]{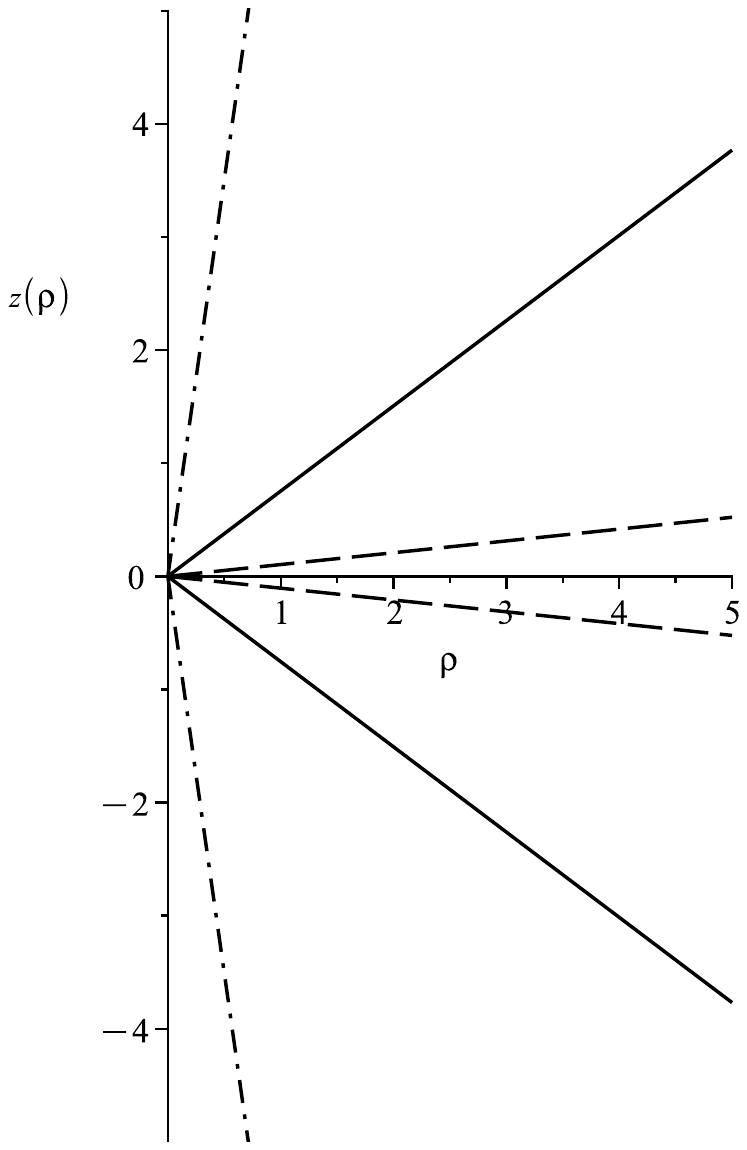}
\caption{The figure shows embedding diagrams of the metric~(\ref{sssss}) for $\beta =-1$. The $\theta$-slices are plotted by ussing Eq.~(\ref{zderho}) for $\theta =0.1$ (dash-dotted line), $\theta =0.6$ (solid line), $\theta = 0.78$ (dashed line).}
\label{conicalbetafijo}
\end{figure}

As an illustrative example, we take the value $\beta=-1$. From Eqs.~(\ref{Cbeta0}) and~(\ref{Cbeta1}), we find that the $\theta$-slices that can be embedded into the ordinary three-dimensional Euclidean space lie within the ranges \(0 < \theta \leq \pi/4\) and \(3\pi/4 \leq \theta < \pi\). This is illustrated in Fig.~\ref{conicalbetafijo}. In this case the angle $\alpha$ varies between $0 < \alpha<\pi/2$ for $0 < \theta \leq \pi/4$, and $-\pi/2 < \alpha<0$ for $3\pi/4 \leq \theta < \pi$. It becomes evident that embeddings for the obtained spacetime do not exist for slices in the range \(\pi/4 < \theta < 3\pi/4\).

\begin{figure}
\centering
\includegraphics[scale=0.3]{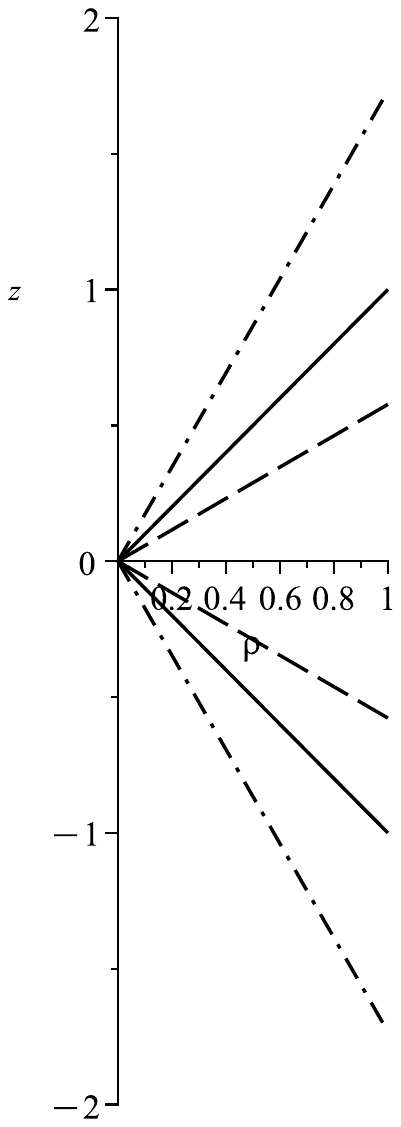}
\caption{The figure displays embedding diagrams of the slice \(\theta = \pi/4\) for the metric~(\ref{sssss}) across various values of the parameter \(\beta\). The embeddings are shown for \(\beta = -1/2\) (dashed line), \(\beta = 0\) (solid line), and \(\beta = 1/2\) (dash-dotted line). Comparing the values of \(z_0(\beta)\) at \(\rho = 1\) using Eq.~(\ref{zetadecero}), we find that \(z_0(-1/2) < z_0(0) < z_0(1/2)\).}
\label{comparandobetass}
\end{figure}  

To more clearly see the effect of the parameter \(\beta\) on the spacetime~(\ref{sssss}) and on the embeddings, we can make use of Eq.~(\ref{zderho}). For this, we may select a specific slice by fixing the spherical polar angle \(\theta = \theta_0\) and the radius \(\rho = \rho_0\). Then, from Eq.~(\ref{zderho}), we have
\begin{eqnarray}
z_0(\beta) = \pm \frac{\left( \frac{1}{1 - \beta} - \sin^2\theta_0 \right)^{\frac{1}{2}}}{\sin\theta_0} \, \rho_0.
\label{zetadecero}
\end{eqnarray}
It is clear from this expression that $z$ depends only on the parameter $\beta$. We compare the behavior of \(z\) with respect to flat spacetime. For \(\beta = 0\), we obtain \(z_0(0) = \pm \frac{\rho_0}{\tan \theta_0}\). Thus, we can conclude that for \(0 < \beta < 1\), \(z_0(\beta) > z_0(0)\), while for \(\beta < 0\), \(z_0(\beta) < z_0(0)\). This situation is illustrated in Fig.~\ref{comparandobetass}, where it can be seen that the cone angle \(\alpha\) decreases as the parameter \(\beta\) increases. In other words, for \(0 < \beta < 1\), the cones become more constricted, while for \(\beta < 0\), the cones appear more open. Therefore, the metric~(\ref{sssss}) globally presents an angular deficit for \(0 < \beta < 1\) and an angular excess for \(\beta < 0\).

\section{Embeddings of Spatial Slices of Schwarzschild-like Wormholes}
We now focus on the embeddings of two-dimensional spatial slices, defined by $t = \text{const}$ and $\theta = \text{const}$, of the metric~(\ref{schwarzschild_like_wormholes}). These slices are described by
\begin{eqnarray}
ds_{\text{slice}}^2 = \frac{dr^2}{\left(1 - \beta\right)\left(1 - \frac{r_0}{r}\right)} + r^2 \sin^2\theta \, d\phi^2. \label{sliceschwarzschild_like_wormholes}
\end{eqnarray}
Unlike the scenario analyzed in the previous section, we now take into consideration the effect of the wormhole throat, located at $r_0$. In such a way, we can now study the deviations from the standard Schwarzschild wormhole arising from the inclusion of the parameter $\beta$ in the metric~(\ref{schwarzschild_like_wormholes}). 

In order to do this the Eqs.~(\ref{rhotheta}) and~(\ref{conicasssssG}) must be considered. Using these equation we may write
\begin{eqnarray}
z(\rho) = \pm \frac{1}{\sin \theta} \int \left(\frac{1}{\left(1-\beta\right)\left(1-\frac{\rho_0}{\rho}\right)}-\sin^2 \theta \right)^{\frac{1}{2}} \, d\rho  \nonumber \\ +C^{\pm}, \,\,\,\,\,\,\,\,\,
\label{ecparametricarho}
\end{eqnarray}
where $\rho_0=r_0 \sin \theta >0$. From this equation, it becomes evident that the condition
\begin{eqnarray}
\frac{1}{\left(1-\beta\right)\left(1-\frac{\rho_0}{\rho}\right)} - \sin^2 \theta \geq 0 
\label{inecuaciongeneral}
\end{eqnarray}
must be satisfied for the integral to exist. In this way, from Eq.~(\ref{inecuaciongeneral}), we can determine the conditions that the parameters must satisfy for the embeddings to exist. For clarity, it is useful to separate these conditions based on whether $\beta$ is positive or negative.

For \(0 \leq \beta < 1\), the inequality~(\ref{inecuaciongeneral}) holds for all values of \(\beta\), allowing all slices in the range \(0 < \theta < \pi\) to be embedded in three-dimensional Euclidean space. These embeddings extend from the throat \(\rho_0\) to infinity.

Now, for the case of $\beta < 0$, we obtain the following inequalities:
\begin{itemize}
\item For the slice $\theta = \pi/2$, we find
\begin{eqnarray}
\rho_0 < \rho \leq \frac{\rho_0 (\beta - 1)}{\beta}.
\label{c1}
\end{eqnarray}
Thus, the embeddings of the equatorial plane extend from the throat to a maximum value of $\frac{\rho_0 (\beta - 1)}{\beta}$.
\item For slices $0<\theta<\pi$ with $\theta \neq \pi/2$,  if $\beta$ satisfies the inequality
\begin{eqnarray}
1-\frac{1}{\sin^2 \theta} \leq \beta <0,
\label{c2}
\end{eqnarray}
then $\rho>\rho_0$, and the embeddings extend from the throat to infinity.
\item For slices $0<\theta<\pi$ with $\theta \neq \pi/2$,  if $\beta$ satisfies the inequality
\begin{eqnarray}
\beta < 1-\frac{1}{\sin^2 \theta}, 
\label{c3}
\end{eqnarray}
then
\begin{eqnarray}
\rho_0 < \rho \leq \frac{\rho_0 (1-\beta) \sin^2 \theta }{(1-\beta) \sin^2 \theta-1},
\label{c4}
\end{eqnarray}
and the embeddings extend from the throat to a maximum value $
\rho_{max}=\frac{\rho_0 (1-\beta) \sin^2 \theta }{(1-\beta) \sin^2 \theta-1}$.
\end{itemize}
As illustrative examples, we now plot several embeddings of slices of the Schwarzschild-like metric. To do this, we use the general form of the function $z(\rho)$, obtained by integrating Eq.~(\ref{ecparametricarho}). The analytic expression for $z(\rho)$ is quite lengthy, so we omit it here.

\subsection{Slices with $\theta=\pi/2$}
\begin{figure}
\centering
\includegraphics[scale=0.3]{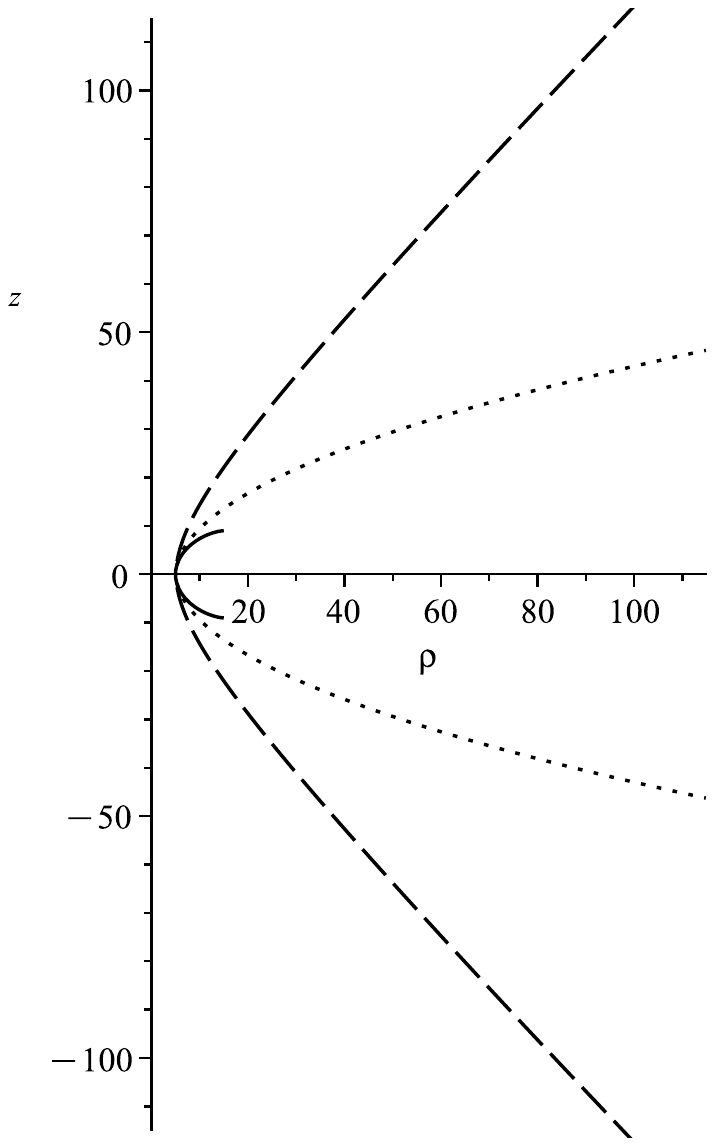}
\caption{The figure displays embedding diagrams of equatorial slice \(\theta = \pi/2\) for the metric~(\ref{schwarzschild_like_wormholes}). The diagrams are plotted for $\rho_0=5$, \(\beta = 1/2\) (dashed line), \(\beta = 0\) (dotted line), and \(\beta = -1/2\) (solid line).}
\label{planoeacuatorial1}
\end{figure}

As mentioned above, embeddings of equatorial plane slices of the metric~(\ref{schwarzschild_like_wormholes}) exist for different values of \( \rho_0 \) and \( \beta \). For \( 0 \leq \beta < 1 \), these embeddings extend from the throat to infinity, while for negative values of \( \beta \), they extend from the throat to a maximum radius determined by constraint~(\ref{c1}). This aspect is illustrated in Fig.~\ref{planoeacuatorial1}, where we have used the values \( \beta = -1/2 \), \( \beta = 0 \), and \( \beta = 1/2 \), with the fixed value \( \rho_0 = 5 \).

From the figure, it is clear that far from the throat, the metric with \( \beta = 1/2 \) exhibits a conical structure, indicating it is not asymptotically flat. For values $\rho \gtrsim 5$, the throat smooths out the conical structure and the conical singularity vanishes. The  metric~(\ref{schwarzschild_like_wormholes})may exhibit asymptotic flatness only when $\beta=0$. For the slice with the negative value \( \beta = -1/2 \), the embedding is possible only from the throat up to \( \rho = 15 \), as defined by inequality~(\ref{c1}).

\subsection{Slices with $\theta \neq \pi/2$}
As established earlier, slices in the range \(0 < \theta < \pi\) can always be embedded in three-dimensional Euclidean space for \(0 \leq \beta < 1\), and these embeddings extend from the throat to infinity. To embed slices in the range \(0 < \theta < \pi\) for negative values of \(\beta\), we must consider Eqs.~(\ref{c2})--(\ref{c4}). 

\begin{figure}
\centering
\includegraphics[scale=0.3]{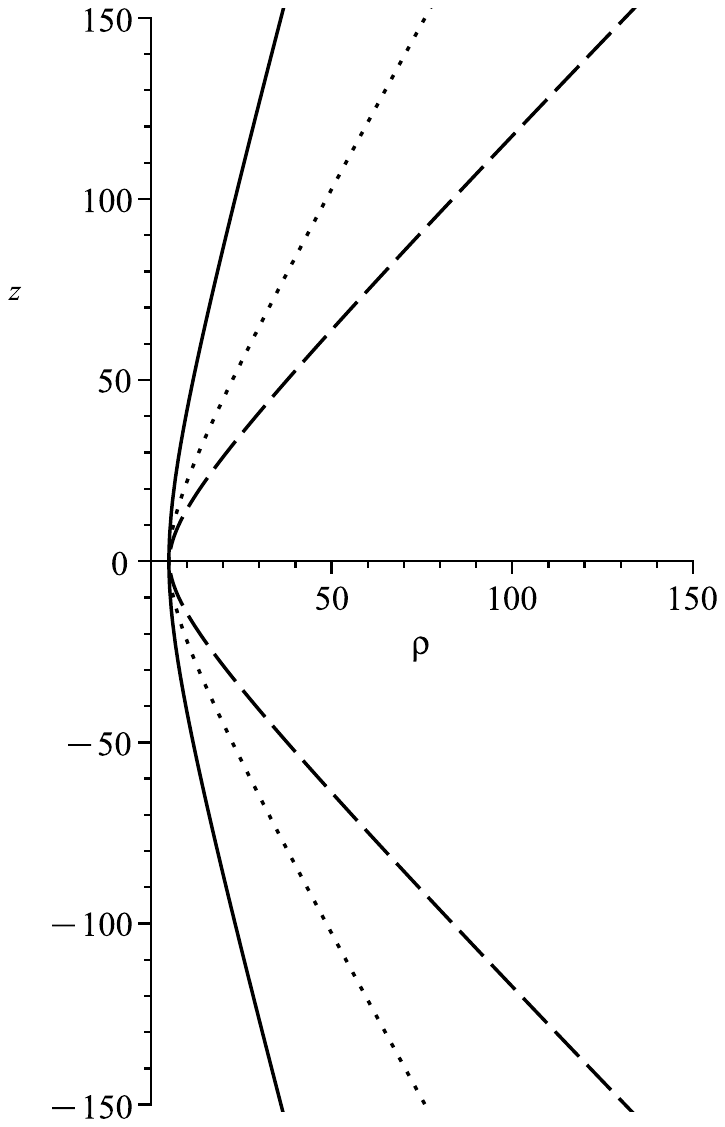}
\caption{The figure displays embedding diagrams of metric~(\ref{schwarzschild_like_wormholes}) with $\beta=1/2$ and $\rho_0=5$. The diagrams are plotted for slices $\theta =\pi/2$ (dashed line) $\theta=\pi/4$ (dotted line) and $\theta =\pi/8$ (solid line).}
\label{beta1215}
\end{figure}

To begin our analysis, we will focus on a specific example within the range $0 \leq \beta < 1$. In Fig.~\ref{beta1215}, embeddings are shown for $\beta = 1/2$ and $\rho_0=5$. For comparison, the slice $\theta = \pi/2$ is also included. As expected, these embeddings extend from the throat $\rho_0$ to infinity.

\begin{figure}
\centering
\includegraphics[scale=0.3]{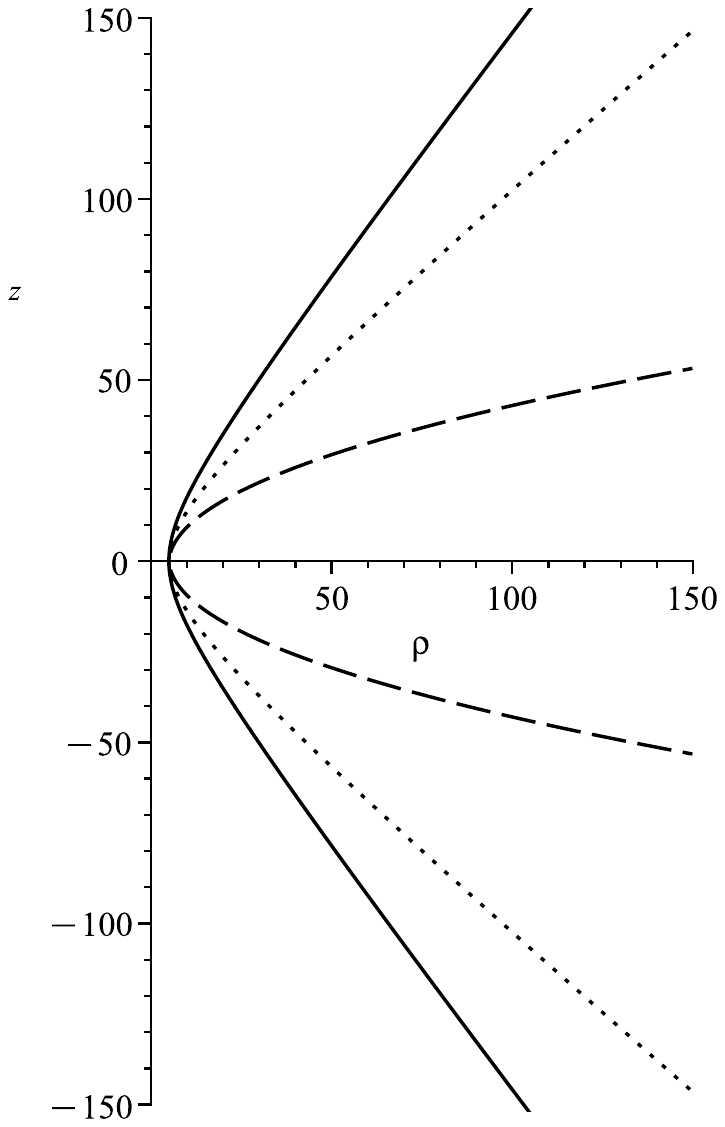}
\caption{The figure displays embedding diagrams of metric~(\ref{schwarzschild_like_wormholes}) with $\beta=-3$ and $\rho_0=5$. The diagrams are plotted for $\theta =\pi/6$ (dashed line) $\theta=\pi/8$ (dotted line) and $\theta =\pi/10$ (solid line).}
\label{betanegativo1215}
\end{figure}

Now, let us consider examples for negative values of $\beta$. As an illustrative case, we explore the specific example with $\beta =-3$. To consider embeddings that extend from the throat to infinity, Eq.~(\ref{c2}) must be taken into account. This equation indicates that slices with $0< \theta \leq \pi/6$ and $5 \pi/6 \leq \theta < \pi$  can be embedded into three-dimensional Euclidean space. In Fig.~\ref{betanegativo1215}, we present the embeddings of metric~(\ref{schwarzschild_like_wormholes}) for $\beta = -3$ and $\rho_0 = 5$, corresponding to the slices $\theta = \pi/6$, $\theta = \pi/8$, and $\theta = \pi/10$. 

\begin{figure}
\centering
\includegraphics[scale=0.3]{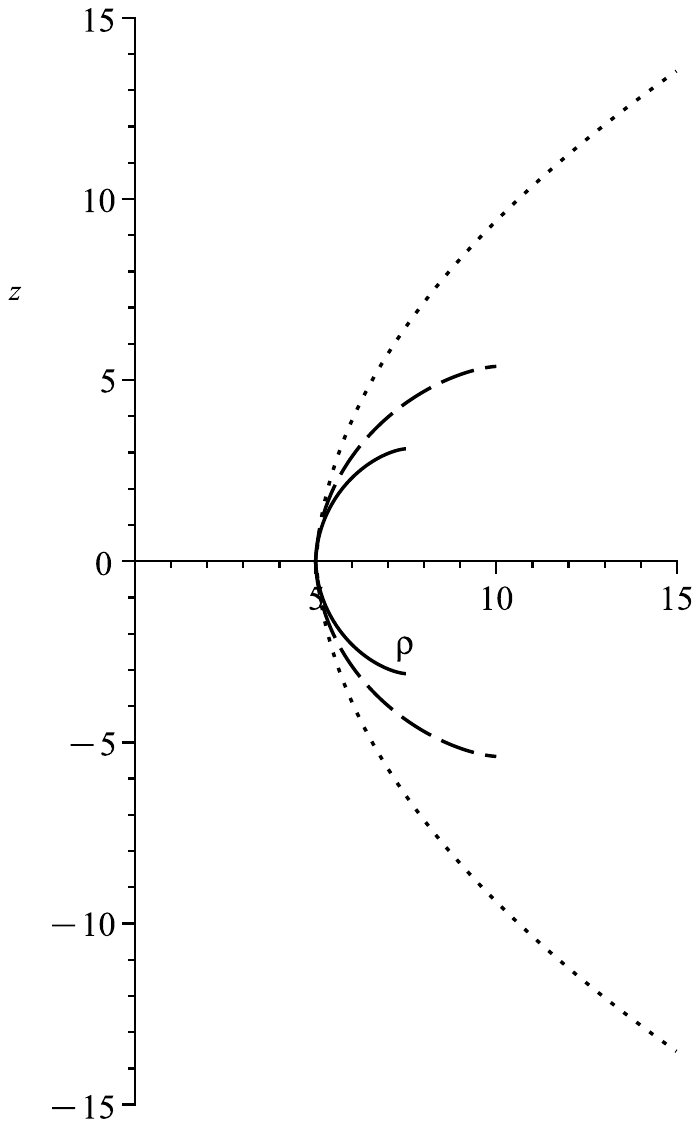}
\caption{The figure displays embedding diagrams of metric~(\ref{schwarzschild_like_wormholes}) with $\beta=-3$ and $\rho_0=5$. The diagrams are plotted for $\theta =\pi/6$ (dotted line) $\theta=\pi/4$ (dashed line) and $\theta =\pi/3$ (solid line).}
\label{betanegativo151215}
\end{figure}

On the other hand, to analyze embeddings that extend from the throat to a maximal radial value, Eq.~(\ref{c3}) must be considered. This equation indicates that slices with $\pi/6 < \theta < 5 \pi/6$ can be embedded into three-dimensional Euclidean space. In Fig.~\ref{betanegativo151215}, we present the embeddings of metric~(\ref{schwarzschild_like_wormholes}) for $\beta = -3$ and $\rho_0 = 5$, corresponding to the slices $\theta = \pi/6$, $\theta = \pi/4$, and $\theta = \pi/3$. In this case, the embedding of the slice $\theta = \pi/6$ extends to infinity, whereas the other two embeddings are limited by the maximal radius: $\rho_{\text{max}} = 10$ for $\theta = \pi/4$ and $\rho_{\text{max}} = 15/2$ for $\theta = \pi/3$.

\section{Conclusions}
This work addresses a fundamental limitation in the traditional visualization of spherically symmetric spacetimes: the inability to embed certain equatorial slices into three-dimensional Euclidean space. While the standard embedding procedure has been successfully applied to the equatorial plane for many well-studied solutions, such as Schwarzschild black holes and Morris–Thorne wormholes, this conventional approach breaks down in the presence of geometries with distinctive topological features, such as an angular excess.

To overcome this limitation, we have developed a generalized isometric embedding procedure that extends the standard approach, traditionally applied to equatorial slices with $t=\text{const}$ and $\theta=\pi/2$ of spherically symmetric spacetimes, to slices with arbitrary polar angles $\theta \neq \pi/2$. This generalized procedure provides a systematic framework for representing two-dimensional slices with $t=\text{const}$ and $\theta=\text{const}$ of a four-dimensional spherically symmetric spacetime within three-dimensional Euclidean space, thereby enabling the analysis and visualization of geometries that cannot be embedded using conventional methods.

We apply the extended embedding procedure to both the conical background metric~(\ref{sssss}) and the Schwarzschild-like wormhole geometry~(\ref{schwarzschild_like_wormholes}). This generalization is necessary because, as noted above, the equatorial slice of the conical background metric cannot be embedded into Euclidean space when the spacetime exhibits an excess solid angle. Consequently, the analysis must be extended to slices with $\theta \neq \pi/2$, particularly for negative values of the parameter $\beta$. At the same time, the method allows us to go beyond the equatorial plane in the case of Schwarzschild-like wormholes, thereby providing a more complete framework for their visualization and analysis. 

In this context, the procedure consists of deriving the embedding equations for both the conical background and the Schwarzschild-like wormhole geometries, and solving them to explicitly characterize the corresponding slices. The discussion examines the key conditions under which the embeddings of specific slices of Schwarzschild-like wormhole solutions can either extend to infinity or terminate at finite radii. In other words, for certain slices, a complete isometric embedding is not possible. These slices are characterized by having negative Gaussian curvature. Effectively, for the two-dimensional slice~(\ref{sliceschwarzschild_like_wormholes}), we find that $2K = R = (\beta - 1)r_0 / r^3$, where $K$ is the Gaussian curvature and $R$ is the scalar curvature. Consequently, for $r_0 \neq 0$, the slice~(\ref{sliceschwarzschild_like_wormholes}) exhibits a negative Gaussian curvature for all values $\beta < 1$. For $r_0 = 0$, we find $K = R = 0$. This result is entirely consistent for $\beta = 0$, as it corresponds to the Euclidean space. However, it is important to emphasize that for $\beta \neq 0$, although the curvature vanishes everywhere for the metric~(\ref{conicalslice}), this conclusion is valid except at $r = 0$, where a conical singularity is present.

Lastly, let us conclude with some remarks about the metric~(\ref{tsssss}). Unlike the background metric~(\ref{sssss}), where the factor $1-\beta$ modifies the radial part (while the angular part retains the standard form of a spherically symmetric metric), in Eq.~(\ref{tsssss}), the factor $1-\beta$ affects the angular part of the metric. Consequently, the surfaces of $r = \text{const}$ are spheres with an effective radius $\sqrt{1-\beta}r$. Now, if we apply the embedding procedure from Sec.~\ref{SectionII} to the slice $t = \text{const}$ and $\theta = \text{const}$ of the metric~(\ref{tsssss}), and comparing it with the metric~(\ref{cylindrical}), we find that
\begin{eqnarray}
\rho = \sqrt{1-\beta} \, r \sin \theta,
\label{z(rho)}
\end{eqnarray}
and therefore the $z$-function is given by
\begin{eqnarray}
z(r) = \pm \sqrt{1 - (1-\beta)\sin^2 \theta} \, r.
\label{zz(r)}
\end{eqnarray}
Combining Eqs.~(\ref{z(rho)}) and~(\ref{zz(r)}), we conclude that the function $z(\rho)$ in cylindrical coordinates is expressed by Eq.~(\ref{zderho}). In this way, all the conclusions presented in Sec.~\ref{SectionIII} also hold for the metric~(\ref{tsssss}).

\begin{acknowledgments}
This work was supported by the Dirección de Investigación y Creación Artística at the Universidad del Bío-Bío through grants No. RE2320220 and GI2310339 (MC).
\end{acknowledgments}

{\bf Data availability} No new data were created or analyzed in this study. Data sharing is not applicable to this article

\end{document}